\documentclass[12pt]{iopart}

\expandafter\let\csname equation*\endcsname\relax
\expandafter\let\csname endequation*\endcsname\relax 

\usepackage{amsmath}
\usepackage{amssymb} 
\usepackage{graphicx}
\usepackage{dcolumn}
\usepackage{bm}
\usepackage{color}
\usepackage{psboxit,epsfig,wrapfig,boxedminipage,enumerate}
\usepackage{nicefrac}

\newcommand{\kt}{k_{\rm B}\hat{T}}
\newcommand{\tb}{\textcolor{black}}
\newcommand{\rij}{{\bf r}_{ij}}
\newcommand{\vecb}[1]{{\rm \bf {#1}}}
\newcommand{\vecbm}[1]{{\bm {#1}}}
\newcommand{\hrij}{{\bf \hat{r}}_{ij}}
\newcommand{\R}{R_{j\vecb{n}}}
\newcommand{\hR}{{\bf \hat{R}}_{j\vecb{n}}}
\def\dblone{\hbox{$1 \hskip -1.2pt\vrule depth 0pt height 1.6ex width
    0.7pt \vrule depth 0pt height 0.3pt width 0.12em$}}
\newcommand{\one}{\dblone}
\newcommand{\MT}{\vecb{M}}

\begin{document}

\title[Ewald sum for hydrodynamic interactions with periodicity in two 
  dimensions]{Ewald sum for hydrodynamic interactions with periodicity in two 
  dimensions}

\author{J Bleibel$^{1,2}$}
\address{$^1$ Max-Planck-Institut f\"ur Intelligente Systeme,
  Heisenbergstr. 3, 70569 Stuttgart, Germany}
\address{$^2$Institut f\"ur theoretische und angewandte Physik,
  Universit\"at Stuttgart, Pfaffenwaldring 57, 70569 Stuttgart, Germany}

\ead{bleibel@mf.mpg.de}

\date{\today}

\begin{abstract} 
  We carry out the Ewald summation for the Rotne--Prager--Yamakawa
  mobility tensor, the Oseen mobility tensor and further variations of both,
  relevant for the hydrodynamic interactions in colloidal suspensions, where
  all interacting particles are within a single plane, i.e., adsorbed at a
  fluid interface or other quasi two--dimensional systems. We use the Poisson
  summation formula for systems periodic in two dimensions and finite in the
  third dimension in order to obtain simple formulae for applications, such as
  molecular dynamics or Brownian dynamics simulations. We show, that for
  such systems, as soon as noise is taken into account, a commonly used
  approximate three--dimensional Ewald summation leads to a spurious system
  size dependence, which may considerably affect the interpretation of
  simulation results and will be cured within our approach. Additionally,
  the resulting formulae are found to be computationally much less expensive
  than the approximate three--dimensional Ewald summation. 
\end{abstract} 


\maketitle



\section{Introduction}
\label{sec:1}
The influence of hydrodynamic interactions (HI) on the dynamics of colloidal
systems (i.e. colloidal suspensions or colloids trapped at an interface) is
subject to ongoing
research~\cite{Rinn:1999,Pesche:2001,Tanaka:2000,Kollmann:2002,Leonardo:2008,Winter:2009,Putz:2010},
both from theoretical and experimental points of view. In many circumstances,
these many body systems can be investigated only with the help of
simulations. For colloids floating in a bulk solvent at low Reynolds number, a
reasonable treatment of HI can be achieved within Stokesian
dynamics~\cite{Brady:1988}. In particular the Oseen or Rotne--Prager--Yamakawa
far-field approximation~\cite{Rotne:1969,Yamakawa:1970}, treating the HI as
pairwise additive interactions, allows an implementation of Hydrodynamic
Interactions within Brownian dynamics computer
simulations~\cite{Brady:1988,Allen:1987}. Since the hydrodynamic interactions
in the bulk decay $\propto 1/r$, where $r$ is the distance between particles
(a similar component is also present in the vicinity of or at
fluid-fluid-interfaces~\cite{Jones:1975,Cichocki:2004}), 
they are considered to be long--ranged and demand special treatment within
simulations. A suitable tool is provided by the Ewald summation of the
Rotne--Prager--Yamakawa mobility
tensor~\cite{Rinn:1999,Pesche:2001,Kollmann:2002,Beenakker:1986}.    

If the colloidal particles in a solvent are not distributed in the full 3D
space, but rather form a thin (mono) layer, the system of interacting colloids
may be considered quasi two--dimensional. This may be realized by either
trapping the particles at an
interface~\cite{Jones:1975,Bleibel:2011,Bleibel:2011EPJ} or by 
placing them in the vicinity of a free or hard
boundary~\cite{Rinn:1999,Kollmann:2002,Cichocki:2004,Zahn:1997}, or by looking
at thin fluid films only~\cite{Leonardo:2008}. Then the question arises how
to treat the hydrodynamic interactions in these quasi two--dimensional systems. 
In the latter case, for colloids in a thin fluid film, an experimental study
revealed, that the two--dimensional form of the Oseen hydrodynamic tensor   
provides a suitable description of the hydrodynamic interactions of this
system~\cite{Leonardo:2008}. However, for colloids
trapped at fluid interfaces~\cite{Bleibel:2011,Bleibel:2011EPJ}, or in the
vicinity of interfaces as in the experiment described in
Refs.~\cite{Rinn:1999,Kollmann:2002}, the situation is more involved. Owing to
the flow fields extending over half the 3D space, the system cannot be
described with a 2D Oseen tensor with its peculiar long--ranged interactions
decaying logarithmically with the interparticle distance. However, also the use
of the full 3D Rotne--Prager--Yamakawa or Oseen tensors seems somewhat
ill--founded, since both do not resemble solutions of the Stokes equation with
respect to the underlying boundary conditions~\cite{Cichocki:2004}. However,
in the case of spherical objects these tensors still might serve as a rough
approximation. As was advocated in ref.~\cite{Jones:1975}, the presence of a
free interface separating two fluid phases, has only little effect on the
diffusion, and thus the mobility, of spherical particles half immersed in both
phases. The Green's function in Stokes flow (Stokeslet) for the
velocity field for a single particle in the presence of a boundary (free
interface or rigid wall) consists of the Oseen tensor, the free (bulk)
solution, plus a mirror term~\cite{Jones:1975,Cichocki:2004,Swan:2007} (method
of images). Therefore, neglecting the latter term while constructing a
solution can be considered as a leading order approximation, where the particles
are assumed to be far from any confining boundary. This has been successfully
applied to simulate the diffusion of particles close to an interface in
experiments~\cite{Rinn:1999,Kollmann:2002}. Alternatively, the quasi 2D
version of the mobility tensor given in Ref.~\cite{Cichocki:2004} for
particles in the vicinity of a free interface may be used.    

For computer simulations using Ewald summation of a 3D mobility
tensor, the question arises how to treat the long ranged part in the  third
($z$-) direction. Within the standard approach as developed by
Beenakker~\cite{Beenakker:1986}, and used in the simulation studies presented
in  Refs.~\cite{Rinn:1999,Kollmann:2002}, one has to assume periodic 
images also in $z$-direction, i.e. the construction of a layered system of many
interfaces. Although easily implemented, such a procedure is computationally
expensive and not necessary from a physical point of view. In view of the
usually implemented 3D Ewald algorithm it is, however, unavoidable.  Since we
are interested in a quasi two--dimensional system which is now extended into the
third dimension, the question arises to which extent the influence of the
artificial periodic images in the $z$--direction disturbs the result of the
summation. Indeed, one could move the periodic images in $z$-direction to large
distances in order to study their influence within the main layer of
particles as function of their distance, as it was done for e.g. Coulomb
interaction (see Ref.~\cite{Yeh:1999} and references therein). Although the
effective velocities for the particles within this procedure converge to the 2D
result, we will show in the following, that as soon as noise is considered,
this approach generates a spurious system size dependence. Additionally, if one
places the periodic images far away from the layer under consideration, this
is even more expensive, since the sums within the usual Ewald formalism have
to be cut off at larger $K$--values in reciprocal space. 
Thus for quasi 2D systems, 3D Ewald summation should be avoided.
However, 2D Ewald summation formulae have been given so far only for the Oseen
tensor in an implicit form~\cite{Pozrikidis:1996}. 
With regard to broader applications, note that the Oseen tensor 
suffers from not being positive definite. This renders its usage problematic
as soon as a noise term 
requiring Cholesky decomposition of the tensor is present. 
Therefore 2D Ewald summation is needed for more suitable mobility tensors.
Our method, as outlined in the following, naturally applies for the
Rotne--Prager--Yamakawa tensor and also in the case of the quasi 2D 
mobility tensor of Ref.~\cite{Cichocki:2004}.     

The paper is organized as follows. In section \ref{sec:2}, we will first
discuss the shortcomings of the three--dimensional Ewald summation for quasi
2D systems. Then we will formally derive the two--dimensional summation
formula for the Rotne--Prager--Yamakawa tensor. As argued above, we
consider this summation of the mobility 
tensor to be the more appropriate approximation than summing up the full 3D
tensor since the result does not suffer from a spurious system--size
dependence due to unphysical images across many additional interfaces and
provides a computationally much cheaper way to incorporate HI within quasi
two--dimensional systems.  
We follow the procedures described in Refs.~\cite{Grzybowski:2000,Porto:2000},
where a lower--dimensional Ewald summation has been developed for
electrostatic and dipole interactions, and derive summation formulae for the
aforementioned mobility tensors with periodicity assumed in two of three
dimensions. 
In section \ref{sec:3}, we demonstrate the 2D Ewald summation procedure
by carrying out simulations of a quasi two--dimensional system. In a first
step, we will show that conventional 3D Ewald summation leads to a divergent
long time diffusion constant, as the system size increases. We then apply our
new summation formulae, and show, that the results are now independent of the
system size. Finally we compare our findings to experimental data, and show
that for the particular system under consideration, a reasonable agreement
with the data can not be achieved using the Rotne--Prager--Yamakawa mobility
tensor. Only upon summing the quasi 2D mobility tensor given by Cichocki and
collaborators, simulations are found to approximate the experimental data.
In view of the latter finding, and since the    
quasi 2D Ewald summation procedure outlined in the following could in
principle be applied to any other geometrical setup or approximation (provided
the system is quasi 2D and terms $\propto 1/r^k$ ($k \ge 1$) are present), 
we have derived and provide explicit 
formulae ready to be used in quasi--2D simulations for the 2D--Ewald sum of 
\begin{enumerate}
\item the Rotne--Prager--Yamakawa tensor (Eqs. (\ref{eq33}-\ref{eq35}))
\item the Oseen tensor (Appendix, Eqs. (\ref{eqCI5}-\ref{eqCI7})) 
\item the quasi 2D mobility tensor of Cichocki et
  al. (Appendix, Eqs. (\ref{eqCI5}-\ref{eqCI7})) 
\item the binary Rotne--Prager tensor (Appendix, Eqs. (\ref{eqC2}-\ref{eqC4}))
\end{enumerate}
Finally, our conclusions are summarized and discussed in section \ref{sec:4}.
 
\section{Ewald summation for quasi two dimensional systems}
\label{sec:2}
Consider a tetragonal lattice with unit cells of volume $L^2\times L^{'}$. The
lattice is periodic in two dimensions (i.e. $x$ and $y$) and finite 
in the third dimension ($z$). Each cell contains $N$ spherical particles,
arranged to form a single layer (monolayer) parallel to the $x-y$ plane.
The force acting on an individual particle $i$ will be denoted by
$\vecb{F}_i$. We assume that no external forces are present, thus the total
force on the particles in the unit cell cancels to zero~\cite{Beenakker:1986}:  
\begin{equation}
  \label{eq1}
  \sum_{i=1}^{N} \vecb{F}_i = 0
\end{equation}
If the particles are surrounded by a solvent, one expects additional
hydrodynamic interactions between the particles. For solvents with low
Reynolds number the motion of the colloidal particles is overdamped and
inertia of the particles may be neglected~\cite{Allen:1987}. Concerning the
implementation of hydrodynamic interactions within computer simulations, this
leads to the use of an position--dependent mobility tensor for the calculation
of the viscous drag of the particles~\cite{Brady:1988,Allen:1987}. One
particular version of this mobility tensor is the Rotne--Prager--Yamakawa
tensor. For the integration of the equations of motion of the colloids, an
effective velocity of each particle has to be calculated via  
\begin{equation}
  \label{eq2}
  \vecb{v}_{i,\rm eff}=\sum_{j=1}^N \MT_{ij}\vecb{F}_j 
\end{equation} 
with the Rotne--Prager--Yamakawa mobility tensor
\addtocounter{equation}{+1}
\begin{align*}
  \label{eq3}
  \MT_{ij} &=(6\pi\eta a)^{-1}\left\{\frac{3}{4}ar^{-1}_{ij}(\one
  +\hrij\hrij)\right.\\ 
  {\tag{\theequation a}}
  &+\left.\frac{1}{2}a^3r^{-3}_{ij}(\one-3\hrij\hrij)\right\}\,,\quad (i\ne
  j\,,\; r \ge 2a)\\
  \label{eq3b}
  \tb{\MT_{ij}} &= \tb{(6\pi\eta a)^{-1}\left\{\left(1-\frac{9}{32}
  \frac{r_{ij}}{a}\right)\one\right.} \\
  {\tag{\theequation \tb{b}}}
  &+\tb{\left.\frac{3}{32}\frac{\rij\rij}{ar_{ij}}\right\}
  \,,\quad (i\ne j\,,\; r < 2a)}\\ 
  {\tag{\theequation c}}
  \MT_{ii}&=(6\pi\eta a)^{-1}\one\,,\quad (i=j)
\end{align*}
and $r_{ij}=|{\bf r}_i-{\bf r}_j|$. The product $\hrij\hrij$ is the outer
product of the normalized vectors $\hrij=\rij/r_{ij}$ and $\one$ denotes the
unity matrix. Since the hydrodynamic interactions are long--ranged $\propto
r^{-1}$, Ewald summation has been suggested to treat the interactions of the
periodic images~\cite{Beenakker:1986}. This leads to a lattice sum 
\begin{equation}
  \label{eq4}
  \vecb{v}_{i,\rm eff}=\sum_{j=1}^N\sum_{\bf n}\,^{\bm \prime}\,
  \MT_{ij}(\rij,{\bf n})\vecb{F}_j 
\end{equation}
where the second sum runs over two dimensional lattice vectors \tb{${\bf
  n}=(n_xL,n_yL)$} with $\vecb{n}\ne 0$ for $\rij= 0$ (indicated by the prime on
the sum) and 
\begin{align}
  \label{eq5}
  \MT_{ij}(\rij,{\bf n}) &=(6\pi\eta a)^{-1}\nonumber\\
  &\times \left\{\left(\frac{3}{4}a\frac{1}{|\rij+{\bf n}|}+\frac{1}{2}a^3
  \frac{1}{|\rij+{\bf n}|^3}\right) \one \right.\nonumber\\
  &+\frac{3}{4} a \frac{1}{|\rij+{\bf n}|^3}(\rij+{\bf n})(\rij+{\bf
    n})\nonumber\\
  &-\left.\frac{3}{2} a^3\frac{1}{|\rij+{\bf n}|^5}(\rij+{\bf n})(\rij+{\bf n})
  \right\}.
\end{align}
\tb{Note that the definition of $\MT_{ij}$ for distances $r<2a$
  (Eq. (\ref{eq3b})), introduced to guarantee the positive definiteness of
  $\MT_{ij}$~\cite{Rotne:1969}, is not relevant for the following. It
  only contributes to the lattice sum for $\vecb{n}=0$ and can be added
  separately.}

\subsection{System size dependence of conventional 3D Ewald summation for
  monolayers} 
Concerning the above lattice sum, we will first consider its
three--dimensional analog and the resulting Ewald summation derived by
Beenakker~\cite{Beenakker:1986}. In order to use it, one has to assume
periodicity in the third dimension, thus the layer of particles is
reproduced also in $z$-direction at distances $n_z
L^{\prime}$~\cite{Rinn:1999,Pesche:2001}. The three--dimensional
lattice sum is split into two sums, one in real and one in reciprocal space,
respectively. We consider the sum in $k$-space ($\vecb{k}$ denotes the
three--dimensional wavevector), 
\begin{equation}
  \label{3dEW1}
  S_k=\frac{1}{L^2L_z} \sum_\vecb{k\ne 0} \sum_{j=1}^N M^{(2)}(\vecb{k}) 
  \vecb{F}_j \cos(\vecb{k}\cdot\rij)
\end{equation}
where the matrix $M^{(2)}(\vecb{k})$ contains the $k$-space part of the
summed mobility tensor (see Eqs. (4) and (6) in ref.~\cite{Beenakker:1986}),
and concentrate on the wavevectors $\vecb{k}$ with $k_x=k_y=0$. Since the
particles are arranged within a single layer, say at $z=0$, the product
$\vecb{k}\cdot\rij$ vanishes for $k_x=k_y=0$ and all pairs of particles. Thus
the cosine above equals one. If we move the layer in 
$z$-direction to large distances, the fundamental mode $k_z^{min}=2\pi/L_z$
approaches zero. For the matrix $M^{(2)}(\vecb{k})=M^{(2)}(k_z)$ then follows:
\begin{align}
  \label{3dEW2}
  \lim_{L_z\to \infty}
  M^{(2)}(k_z)&=\lim_{L_z\to \infty}
  \left(\one-\frac{\vecb{k}\vecb{k}}{k^2}\right) (a-\frac{1}{3}a^3k_z^3)
  \nonumber\\ 
  &\times \left(1+ \frac{k_z^2}{4\alpha^2}+\frac{k_z^4}{8\alpha^4}\right)
  \frac{6\pi}{k_z^2} \exp\left(-\frac{k_z^2}{4\alpha^2}\right)
  \nonumber\\ 
  &\approx \lim_{k_z\to 0} \one\left(\frac{6\pi a}{k_z^2}+\frac{6\pi a}{4
    \alpha^2}-\frac{1}{3}6\pi a^3\right)\nonumber\\
  &-\frac{\vecb{k}\vecb{k}}{k^2}\left(\frac{6\pi a}{k_z^2}+\frac{6\pi a}{4
    \alpha^2}-\frac{1}{3}6\pi a^3\right)
\end{align}
Since the matrix $\vecb{k}\vecb{k}/k^2$ contains only one nonzero element,
($(\vecb{k}\vecb{k}/k^2)_{zz}=1$) the diagonal elements $M^{(2)}_{xx}$ and
$M^{(2)}_{yy}$ of the matrix diverge as $L_z$ increases. 
The value of these matrix elements does not depend on any dynamical
variable, it is constant and determined by the choice of system size $L_z$
and particle radius $a$. Therefore, after carrying out the 3D Ewald
summation, the summed mobility matrix $\MT_{ij}^*(\rij)$ consists of a
dynamical part, and a constant part depending only on system parameters and
diverging as $L_z$ increases \tb{($M^{(1)}(\rij,{\bf n})$ denotes the spatial
  lattice sum of the mobility matrix, c.f. Eq. (5) in
  ref.~\cite{Beenakker:1986})}:  
\begin{align}
  \label{3dEW3}
  \MT_{ij}^*(\rij) &= \sum_{\bf n}\,^{\bm \prime}\,
  \MT_{ij}(\rij,{\bf n})  \nonumber\\ 
  &=(6\pi\eta a)^{-1}\left( \one\, \delta_{ij}+ (1-\delta_{ij}) \left\{
  \sum_{\bf n}\,^{\bm \prime}\, M^{(1)}(\rij,{\bf n})\right.\right. \nonumber\\
  &+ \frac{1}{L^2L_z} 
  \sum_{\vecb{k}\ne 0 \atop k_x \ne 0 \,\lor \,k_y \ne 0} M^{(2)}(\vecb{k}) 
  \cos(\vecb{k}\cdot\rij) 
  +\underbrace{\frac{1}{L^2L_z}\sum_{k_z \ne 0 \atop k_x=k_y=0}
    M^{(2)}(k_z)}_{= const.}\left.\left.\vphantom{\sum_n}\right\}\right)
\end{align}
$\MT_{ij}^*(\rij)$ is in fact a $3N\times3N$ matrix, consisting
of a set of $3\times3$ submatrices $D_{ij}(\rij)$ for each pair of particles.
After the 3D Ewald summation it contains a diverging constant in the
$xx-$ and $yy-$ diagonal elements of all off-diagonal submatrices
$D_{ij,i\ne j}(\rij)$. 
Upon summing over the forces on all particles this constant part does not
contribute, since we assumed a zero net force (see Eqs. (\ref{eq1}) and
(\ref{eq4})). \tb{If the latter requirement is relaxed, i.e. in order to study
  sedimentation, for 3D suspensions it is necessary to include the backflow
  of the solvent as a pressure gradient in order to achieve a cancellation of
  the singular terms arising from $\vecb{k}=0$~\cite{Brady:1988}. However,
  this would not cure the system size dependence in this special case, since
  the divergence with $L_z$ does not require a vanishing wavevector. It is
  rather an artefact of the application of a summation technique appropriate for
  three--dimensional periodic systems only.}     
For the sum in Eq. (\ref{3dEW1}) the divergence \tb{with $L_z$} is therefore
relevant \tb{for non--zero net forces only}. However, the summed mobility
matrix also enters the calculation of the correlated noise~\cite{Allen:1987},  
\begin{equation}
  \langle \vecb{r}_i(t),\vecb{r}_j(t+\Delta t)\rangle =
  2\MT_{ij}^*(\rij)\Delta t\
\end{equation}
thus implying a diverging width  of the correlator. Within the usual Ermak
algorithm, one has to calculate the Cholesky decomposition of
$\MT_{ij}^*(\rij)$ in order to compute $3N$ correlated random numbers
for the random displacement of the particles. Therefore, if
$\MT_{ij}^*(\rij)$ contains a diverging part, its ``square root''
matrix $\vecb{\sigma}_{ij}^*(\rij)$ with $\vecb{\sigma}_{ij}^*
\vecb{\sigma}_{ij}^{*T}=\MT_{ij}^*$ will also diverge
and the random displacement of the particles may become arbitrarily large.
A possible way out could be to simply subtract the divergent part, or to cut
off the sum at a certain value, however, the latter would introduce an
additional parameter whereas in the first case, we are not aware of a
consistent argument, why the distances of the additional layers introduced in
$z-$direction, should not matter. Therefore we consider it more appropriate
to avoid this scenario by considering the system to be genuinely
two--dimensional from the beginning.     

\tb{It is interesting to compare the setup of a monolayer of particles to
  another relevant case: particles confined between two parallel walls,
  rendering a three dimensional distribution of particles with finite
  extent in one dimension (see
  e.g. ref.~\cite{Hernandez:2006,Hernandez:2007L,Swan:2011,Zhang:2012}). A
  monolayer of colloids may be considered a limiting case of the more general
  configuration of a confined suspension of particles. However, there are also
  important differences: Within the confined geometry, particles are not only
  restricted to the slit by the walls, the different boundary conditions of
  the walls compared to the unbound fluid also alter the hydrodynamics. 
  The distribution of particles is finite in one dimension, however,
  depending on the width of the slit, particles are able to move in the third
  direction also. The hydrodynamic interactions then also depend on the
  third spatial coordinate, and the 3D Ewald summation would therefore not
  diverge as in the case of a monolayer. The slit geometry has been dealt with
  in detail in ref.~\cite{Hernandez:2006}, where the hydrodynamic interactions
  were included on the basis of a two dimensional Fourier series for the
  Green's function of the Stokes equation with the corresponding no--slip
  boundary conditions. The method has been generalized to arbitrary domains in
  ref.~\cite{Hernandez:2007L} as a new method for the computation of the
  hydrodynamical interactions for confined geometries, the so--called general
  geometry Ewald--like method (GGEM). The latter method relies on the
  separation of forces into local and long--ranged
  parts~\cite{Hernandez:2007L,Swan:2011} just as in conventional
  Ewald--summation in electrostatics. Instead of solving the Stoke equation in
  order to obtain the Green's function, our approach is rather to use an
  existing 3D solution in terms of well known bulk mobility tensors and apply
  them to a two--dimensional monolayer of particles immersed in an infinite 3D
  medium. In the following, we first keep the dependence on the 
  non--periodic, third spatial coordinate throughout the derivation. Only for
  the final formulas, we then consider the limit of vanishing distances
  between the particles in the third direction, i.e. the spatial configuration
  of a monolayer. However, upon dropping this requirement and after some
  straightforward calculations, summation formulae could also be obtained for
  a three--dimensional configuration of particles, with finite extent in on
  dimension.      
 } 
   
\subsection{Two--dimensional Ewald summation}
  In the following we will now derive the result for the Ewald summation
  with periodicity in only two of three dimensions, which will cure the above
  spurious dependence on the system size. According to the procedure
  described in ref.~\cite{Grzybowski:2000}, we introduce: 
\begin{equation}
  \label{eq6}
  \Phi(\rij)=\sum_\vecb{n}\frac{1}{|\rij+\vecb{n}|}\,,\quad(\rij\ne 0);
\end{equation} 
\begin{equation}
  \label{eq7}
  \Psi(\rij)=\sum_\vecb{n}\frac{1}{|\rij+\vecb{n}|^3}\,,\quad(\rij\ne 0);
\end{equation}
\begin{equation}
  \label{eq8}
  \Theta(\rij,\vecbm{\xi})=\sum_\vecb{n}
  \frac{\exp(-i\vecbm{\xi}(\rij+\vecb{n}))}{|\rij+\vecb{n}|^3} 
  \,,\quad(\rij\ne 0); 
\end{equation}
\begin{equation}
  \label{eq9}
  \chi(\rij,\vecbm{\xi})=\sum_\vecb{n}
  \frac{\exp(-i\vecbm{\xi}(\rij+\vecb{n}))}{|\rij+\vecb{n}|^5} 
  \,,\quad(\rij\ne 0);
\end{equation}
and denote the sums for $\rij = 0$ and $\vecb{n}\ne 0$ by $\Phi_0$,
$\Psi_0$, $\Theta_0$ and $\chi_0$ respectively:
\begin{equation}
  \label{eq6a}
  \Phi_0=\sum_\vecb{n}\frac{1}{|\vecb{n}|}\,,\quad(\vecb{n}\ne 0);
\end{equation} 
\begin{equation}
  \label{eq7a}
  \Psi_0=\sum_\vecb{n}\frac{1}{|\vecb{n}|^3}\,,\quad(\vecb{n}\ne 0);
\end{equation}
\begin{equation}
  \label{eq8a}
  \Theta_0(\vecbm{\xi})=\sum_\vecb{n}
  \frac{\exp(-i\vecbm{\xi}(\vecb{n}))}{|\vecb{n}|^3} 
  \,,\quad(\vecb{n}\ne 0); 
\end{equation}
\begin{equation}
  \label{eq9a}
  \chi_0(\vecbm{\xi})=\sum_\vecb{n}
  \frac{\exp(-i\vecbm{\xi}(\vecb{n}))}{|\vecb{n}|^5} 
  \,,\quad(\vecb{n}\ne 0).
\end{equation}
Using these definitions, Eq. (\ref{eq4}) can be written as
\begin{align}
  \label{eq10}
  &(6\pi\eta a)\vecb{v}_{i,\rm eff}=\vecb{F}_i+\sum_{j=1}^N\left\{\left(
  \frac{3}{4}a\Phi(\rij)\right.\right.\nonumber\\
  &+\left.\frac{3}{4}a\Phi_0+ \frac{1}{2}a^3\Psi(\rij)
  +\frac{1}{2}a^3\Psi_0\right) \one \nonumber\\  
  &-\left.\frac{3}{4}a\nabla_{\bm \xi}\nabla_{\bm \xi}\Theta(\rij,
  \vecbm{\xi})\right|_{\vecbm{\xi}=0}-\left. 
  \frac{3}{4}a\nabla_{\bm \xi}\nabla_{\bm \xi}\Theta_0(\vecbm{\xi})
  \right|_{\vecbm{\xi}=0} \nonumber\\
  &+\left.\left.\frac{3}{2}a^3\nabla_{\bm \xi}\nabla_{\bm \xi}
  \chi(\rij,\vecbm{\xi})\right|_{\vecbm{\xi}=0} +\left. \frac{3}{2}a^3\nabla_{\bm
    \xi}\nabla_{\bm \xi} \chi_0(\vecbm{\xi}) \right|_{\vecbm{\xi}=0} \right\}
  \vecb{F}_j 
\end{align}
where the $\nabla_{\bm \xi}$ denotes the gradient with respect to ${\bm \xi}$
and is evaluated for ${\bm \xi}=0$~\cite{Grzybowski:2000}. Note that the
gradients appear with the opposite sign to compensate for the $i^2$ due to
differentiation of Eq. (\ref{eq8}) and Eq. (\ref{eq9}).  

The sums appearing in Eqs. (\ref{eq6})-(\ref{eq9}) will
now each be transformed into two rapid converging sums in real and reciprocal
space. We repeat the formalism described in Ref.~\cite{Grzybowski:2000}, in
detail for the first component $\Phi(\rij)$ and give the results for
the other parts.  For this purpose we use the definition of the Gamma function 
\begin{equation}
  \label{eq11}
  \frac{1}{r^{2s}}=\frac{1}{\Gamma(s)}\int_0^{\infty} t^{s-1}e^{-r^2t} dt
\end{equation}
and Poisson's summation formula in two dimensions for a sum of Gaussians
\begin{equation}
  \label{eq12}
  \sum_\vecb{n} \exp(-|\vecbm{\rho}+\vecb{n}|^2t)= \frac{\pi}{L^2t}
  \sum_\vecb{K} \exp(i\vecb{K}\vecbm{\rho}) \exp(-\frac{K^2}{4t})
\end{equation}
where $\vecbm{\rho}=(x,y)$ and $\vecb{K}=(k_x,k_y)$ are two--dimensional
vectors in real and reciprocal space respectively. Note that the
three--dimensional version of this summation formula differs only be an
additional factor of $\sqrt{\pi}/Lt^{-1/2}$ on the right hand side, stemming
from the underlying Fourier transformation of the sum of Gaussians. Using the 
3D summation formula instead, e.g. in order to derive the original result of
Beenakker for the 3D Rotne--Prager tensor~\cite{Beenakker:1986}, would
therefore not change the general outline of the calculation, but lead to
different integrals in the following.      

Inserting Eq. (\ref{eq11}) for $s=1/2$ into Eq. (\ref{eq6})
leads to~\cite{Grzybowski:2000} 
\begin{equation}
  \label{eq13}
  \Phi=\sum_\vecb{n} \frac{1}{\Gamma(1/2)} \int_0^{\infty} t^{-1/2}
  \exp(-|\rij+{\bf n}|^2t) dt
\end{equation}
The integral may be split up, by introducing a convergence
factor $\alpha$~\cite{Grzybowski:2000} (still to be determined) 
\begin{align}
  \label{eq14}
  \Phi&=\sum_\vecb{n} \frac{1}{\sqrt{\pi}} \int_{\alpha^2}^{\infty} t^{-1/2}
  \exp(-|\rij+{\bf n}|^2t) dt\nonumber\\
  &+\sum_\vecb{n} \frac{1}{\sqrt{\pi}}
  \int_0^{\alpha^2} t^{-1/2} \exp(-|\rij+\vecb{n}|^2t) dt
\end{align}
The first integral can be evaluated, for the second we apply the summation
formula Eq. (\ref{eq12}):
\begin{align}
  \label{eq15}
  \Phi&=\sum_\vecb{n}\frac{{\rm erfc}(\alpha|\rij+{\bf n}|)} {|\rij+{\bf
      n}|}\nonumber\\ 
  &+\frac{\sqrt{\pi}}{L^2} \int_0^{\alpha^2} t^{-3/2} \sum_\vecb{K}
  \exp(i\vecb{K}\vecbm{\rho}_{ij})  \exp(-\frac{K^2}{4t}-z_{ij}^2t) dt
\end{align}
where ${\rm erfc}(x)=1-{\rm erf}(x)$ is the complementary error function and
$\rij=(\vecbm{\rho}_{ij},z_{ij})$. The integral will contain a singularity for
$K=0$. This term deserves a special treatment, therfore, we separate the term
with $K=0$ from the sum to isolate the singularity~\cite{Grzybowski:2000}. As
will be seen later on we can use the absence of external forces
(Eq. (\ref{eq1})) to get rid off all singular and constant
terms~\cite{Beenakker:1986,Grzybowski:2000}.   
\begin{align}
  \label{eq16}
  \Phi&=\sum_\vecb{n}\frac{{\rm erfc}(\alpha|\rij+{\bf n}|)} {|\rij+{\bf
      n}|}+\frac{\sqrt{\pi}}{L^2} \int_0^{\alpha^2} t^{-3/2} \exp(-z_{ij}^2t) dt
  \nonumber\\ 
  &+\frac{\sqrt{\pi}}{L^2} \sum_\vecb{K\ne 0} \exp(i\vecb{K}\vecbm{\rho}_{ij})
  \int_0^{\alpha^2} t^{-3/2} \exp(-\frac{K^2}{4t}-z_{ij}^2t) dt
\end{align}
The first integral evaluates to
\begin{align}
  \label{eq17}
  &\int_0^{\alpha^2} t^{-3/2} \exp(-z_{ij}^2t) dt =\nonumber\\
  &-2\sqrt{\pi}z_{ij}{\rm erf}(\alpha
  z_{ij})-\frac{2}{\alpha}e^{-\alpha^2z_{ij}^2}
  +\lim_{t\to 0^+} \frac{2e^{-z_{ij}^2t}}{\sqrt{t}}
\end{align}
while the second integral can be performed using the substitution
$u^2=1/t$~\cite{Grzybowski:2000}:
\begin{align}
  \label{eq18}
  &\int_0^{\alpha^2} t^{-3/2}  \exp(-\frac{K^2}{4t}-z_{ij}^2t) dt = \nonumber \\
  &\frac{\sqrt{\pi}}{K}\left[e^{-Kz_{ij}}{\rm erfc}(\frac{K}{2\alpha}-\alpha
    z_{ij}) +e^{Kz_{ij}}{\rm erfc}(\frac{K}{2\alpha}+\alpha z_{ij})\right]
\end{align}  
Putting all the pieces together, one ends up with
\begin{align}
  \label{eq19}
  &\Phi(\rij)=\sum_\vecb{n}\frac{{\rm erfc}(\alpha|\rij+{\bf n}|)} {|\rij+{\bf
      n}|} \nonumber\\
  &+\frac{\pi}{L^2}\sum_{K \ne 0} \frac{\exp(i\vecb{K}\vecbm{\rho}_{ij})}{K}
  \left[e^{-Kz_{ij}}{\rm erfc}(\frac{K}{2\alpha}-\alpha z_{ij})
    \right.\nonumber\\    
  &+\left.e^{Kz_{ij}}{\rm erfc}(\frac{K}{2\alpha}+ \alpha
    z_{ij})\right]\nonumber \\
  &-\frac{2\sqrt{\pi}}{L^2}\left[\sqrt{\pi}z_{ij}{\rm erf}(\alpha z_{ij})+
  \frac{1}{\alpha}e^{-\alpha^2z_{ij}^2} 
  -\lim_{t\to 0^+} \frac{e^{-z_{ij}^2t}}{\sqrt{t}}\right]
\end{align}
Accordingly, for $\Phi_0$ (Eq. (\ref{eq6a})) one finds
\begin{align}
  \label{eq20}
  \Phi_0&=\sum_\vecb{n}\frac{{\rm erfc}(\alpha|{\bf n}|)} {|{\bf n}|}
  +\frac{\pi}{L^2}\sum_{K \ne 0} \frac{2}{K}
  {\rm erfc}(\frac{K}{2\alpha})\nonumber \\
  &-\frac{2\sqrt{\pi}}{L^2\alpha}+ \frac{2\sqrt{\pi}}{L^2} \lim_{t\to 0^+}
  \frac{1}{\sqrt{t}}-\frac{2\alpha}{\sqrt{\pi}}
\end{align}
where the term for $\vecb{n}=0$ had to be inserted in the sum of the integral
on $[0,\alpha^2]$ and subtracted separately in order to apply the Poisson
summation formula~\cite{Grzybowski:2000}.

For the terms $\Psi(\rij)$ and $\Psi_0$ the procedure is similar, so we just
give the results. Note, that no singular terms will appear while
performing the required integrations. Use Eq. (\ref{eq11}) with $s=3/2$ leads
to  
\begin{align}
  \label{eq21}
  \Psi&(\rij)=\frac{2}{\sqrt{\pi}} \sum_\vecb{n} \int_{\alpha^2}^{\infty} t^{1/2}
  \exp(-|\rij+{\bf n}|^2t) dt\nonumber\\
  &+\frac{2\sqrt{\pi}}{L^2} \sum_\vecb{K} \exp(i\vecb{K}\vecbm{\rho}_{ij}) 
  \int_0^{\alpha^2} t^{-1/2}  \exp(-\frac{K^2}{4t}-z_{ij}^2t) dt
\end{align}
and
\begin{align}
  \label{eq22}
  \Psi_0&=\frac{2}{\sqrt{\pi}} \sum_\vecb{n} \int_{\alpha^2}^{\infty} t^{1/2}
  \exp(-|{\bf n}|^2t) dt\nonumber\\
  &+\frac{2\sqrt{\pi}}{L^2} \sum_\vecb{K} \int_0^{\alpha^2} t^{-1/2} 
  \exp(-\frac{K^2}{4t}) dt -\frac{4\alpha^3}{3\sqrt{\pi}}
\end{align}

The $\Theta$ and $\chi$ terms require a different form of the Poisson
summation formula~\cite{Grzybowski:2000}:
\begin{align}
  \label{eq23}
  \sum_\vecb{n} \exp(&-|\vecbm{\rho}+\vecb{n}|^2t-i\vecbm{\xi}
  (\vecbm{\rho}+\vecb{n}))\nonumber\\
  &=\frac{\pi}{L^2t} \sum_\vecb{K} \exp(i\vecb{K}\vecbm{\rho}) 
  \exp(-\frac{|\vecb{K}+\vecbm{\xi}|^2}{4t})
\end{align}
which takes the additional ${\bm \xi}$-dependence into account. 
Using Eq. (\ref{eq11}) with $s=3/2$, Eq. (\ref{eq8}) can be written in the
following form:
\begin{align}
  \label{eq24}
  \Theta(\rij,\vecbm{\xi})&=\frac{2}{\sqrt{\pi}} \sum_\vecb{n}
  \exp(-i\vecbm{\xi}(\rij+\vecb{n}))\nonumber \\ 
  &\times \int_0^{\infty} t^{1/2} \exp(-|\rij+\vecb{n}|^2t) dt
\end{align} 
Again, one splits up the integral and applies Eq. (\ref{eq23}) in the second
integral. Thus
\begin{align}
  \label{eq25}
  \Theta(\rij,\vecbm{\xi})&=\frac{2}{\sqrt{\pi}} \sum_\vecb{n}
  \exp(-i\vecbm{\xi}(\rij+\vecb{n}))\nonumber \\ 
  &\times \int_{\alpha^2}^{\infty} t^{1/2} \exp(-|\rij+\vecb{n}|^2t)dt \nonumber\\
  &+\frac{2\sqrt{\pi}}{L^2} \sum_{\vecb{K}} \exp(i\vecb{K}\vecbm{\rho}_{ij}-
  i\xi_zz_{ij})\nonumber\\
  &\times \int_0^{\alpha^2} t^{-1/2}
  \exp(-\frac{|\vecb{K}+\vecbm{\xi}_{\rho}|^2}{4t}-z_{ij}^2t)dt  
\end{align}
where $\vecbm{\xi}_{\rho}$ denotes the two--dimensional $x$-- and $y$--part of
$\vecbm{\xi}$,  and accordingly for $\Theta_0(\xi)$
\begin{align}
  \label{eq26}
  \Theta_0&(\vecbm{\xi})=\frac{2}{\sqrt{\pi}} \sum_\vecb{n}
  \exp(-i\vecbm{\xi}(\vecb{n})) \int_{\alpha^2}^{\infty} t^{1/2}
  \exp(-|\vecb{n}|^2t)dt \nonumber\\ 
  &+\frac{2\sqrt{\pi}}{L^2} \sum_{\vecb{K}} \int_0^{\alpha^2} t^{-1/2}
  \exp(-\frac{|\vecb{K}+\vecbm{\xi}_{\rho}|^2}{4t})dt
  -\frac{4\alpha^3}{3\sqrt{\pi}}   
\end{align}
The calculation for $\chi(\rij,\vecbm{\xi})$ is carried out with $s=5/2$ and
after some short manipulations one finds
\begin{align}
  \label{eq27}
  \chi(\rij,\vecbm{\xi})&=\frac{4}{3\sqrt{\pi}} \sum_\vecb{n}
  \exp(-i\vecbm{\xi}(\rij+\vecb{n}))\nonumber \\ 
  &\times \int_{\alpha^2}^{\infty} t^{3/2} \exp(-|\rij+\vecb{n}|^2t)dt \nonumber\\
  &+\frac{4\sqrt{\pi}}{3L^2} \sum_{\vecb{K}} \exp(i\vecb{K}\vecbm{\rho}_{ij}-
  i\xi_zz_{ij})\nonumber\\
  &\times \int_0^{\alpha^2} t^{1/2}
  \exp(-\frac{|\vecb{K}+\vecbm{\xi}_{\rho}|^2}{4t}-z_{ij}^2t)dt  
\end{align}
\begin{align}
  \label{eq28}
  \chi_0&(\vecbm{\xi})=\frac{4}{3\sqrt{\pi}} \sum_\vecb{n}
  \exp(-i\vecbm{\xi}(\vecb{n})) \int_{\alpha^2}^{\infty} t^{3/2}
  \exp(-|\vecb{n}|^2t)dt \nonumber\\ 
  &+\frac{4\sqrt{\pi}}{3L^2} \sum_{\vecb{K}} \int_0^{\alpha^2} t^{1/2}
  \exp(-\frac{|\vecb{K}+\vecbm{\xi}_{\rho}|^2}{4t})dt
  -\frac{8\alpha^3}{15\sqrt{\pi}}   
\end{align}

The calculation of the components for $\rij=0$, i.e. $\Phi_0$, $\Psi_0$,
$\Theta_0$ and $\chi_0$ (see Eqs. (\ref{eq6a}-\ref{eq9a})) is only needed to
extract the constant term arising from the constraint $K\ne 0$. This term only
contributes for $i=j$, since it corresponds to the additional part summed up
for $\rij=0$ and $n=0$. Therefore it is placed outside the sum over $j$. The
remaining part can be absorbed into the main formulae by allowing
$\rij=0$. Note that for $\Theta$ and $\chi$ this constant term does not
contribute, since it doesn't depend on ${\bm \xi}$ and only the matrix
elements with gradients are used. A similar singularity as in the integration
(Eq. (\ref{eq17})) will appear when integrating $\nabla_{\bm \xi}\nabla_{\bm
  \xi}\Theta$ (see also Eq. (\ref{A1}) in the appendix).  

We are now left with expressions for $\Phi$, $\Psi$, $\Theta$ and $\chi$ as
sums in real and reciprocal space. It remains to evaluate the gradients with
respect to ${\bm \xi}$, perform the remaining integrals and collect all parts
from Eqs. (\ref{eq19}-\ref{eq22}) and Eqs. (\ref{eq25}-\ref{eq28}) in order to
insert them into Eq. (\ref{eq10}). We list all components and integrals in the
appendix. To obtain the final formula, we take the limit 
$z_{ij}\to 0$ for all pairs of particles, i.e. all particles remain in a plane
parallel to the ($x-y$)--plane. We introduce the following definitions: 
\begin{equation}
  \label{eq29}
  \R=|\rij+\vecb{n}|\,,\quad \hR=\frac{(\rij+\vecb{n})}{|\rij+\vecb{n}|}\,,
  \quad \vecb{\hat{K}}=\frac{\vecb{K}}{K}.
\end{equation}
Since we are only interested in the limit $z_{ij}=0$, we can assume without
loss of generality that $z=0$ for all particles, i.e. $\rij=\vecbm{\rho}_{ij}$,
and thus consider $\hR$, $\vecb{v}_{i,\rm eff}$ and $\vecb{F}_i$ to be
two--dimensional only from now on. Before writing down the sum of all
components  
we note that several terms cancel and, additionally, all
divergent and constant parts stemming from Eqs. (\ref{eq19}-\ref{eq22}) and
(\ref{eq25}-\ref{eq28}) do not contribute to the 
final result. \tb{In the case of a net zero force on the particles in the
  system, (see Eq. (\ref{eq1})), any product with a (diverging) constant also
  vanishes. As discussed in Ref.~\cite{Brady:1988}, the cancellation of the
  singular terms stemming form $K=0$ can be achieved, even in the case when
  the average force on the particles is not zero. To this end, one has to
  introduce a pressure gradient representing the backflow of the fluid. The
  relevant physical quantities are then the velocities relative to the
  backflow of the fluid~\cite{Brady:1988}.} 
Taking all similar parts together, one then arrives at a rather compact
result:  
\begin{align}
  \label{eq32}
  &(6\pi\eta a)\vecb{v}_{i,\rm eff}=\vecb{F}_i
  -\frac{a\alpha}{\sqrt{\pi}}
  \left(\frac{3}{2}+\frac{2}{3}a^2\alpha^2\right)\vecb{F}_i\nonumber\\
  &+\sum_{j=1}^N\left\{\sum_\vecb{n}\,^{\bm \prime}\,\left( \frac{3}{4}a\frac{{\rm
      erfc}(\alpha\R)}{\R}\left[\one+ \hR\hR\right] \right.\right.\nonumber\\
  &+\left.\left[\frac{1}{2}a^3\frac{{\rm erfc}(\alpha\R)}{\R^3}
    +\frac{a^3\alpha}{\sqrt{\pi}}
    \frac{e^{-\alpha^2\R^2}}{\R^2}\right] \left[\one-3\hR\hR\right]
  \right. \nonumber\\ 
  &\left. +\frac{a\alpha}{\sqrt{\pi}}
  e^{-\alpha^2\R^2} \left[\frac{3}{2}-2a^2\alpha^2\right]\hR\hR\right)\nonumber\\
  &+\left.\sum_{\vecb{K}\ne 0} \cos(\vecb{K}\rij)\left(
  \frac{3a\pi}{L^2K} {\rm erfc}\left(\frac{K}{2\alpha}
  \right)\right.\right.\nonumber\\  
  &\times\left[\one-(\frac{1}{2}-\frac{1}{3}a^2K^2)\vecb{\hat{K}}
    \vecb{\hat{K}}\right]\left.\left.  
  -\frac{3a\sqrt{\pi}}{2L^2\alpha} e^{-\frac{K^2}{4\alpha^2}}
      \vecb{\hat{K}}\vecb{\hat{K}}\right)\right\}\vecb{F}_j  
\end{align}

If we cast this result into a similar form as in Ref.~\cite{Beenakker:1986}, 
the final formula reads:
\begin{align}
  \label{eq33}
  (6\pi\eta a)&\vecb{v}_{i,\rm eff}=\sum_{j=1}^N \sum_\vecb{n}\,^{\bm \prime}
  M^{(1)}(\vecb{R}_{j,\vecb{n}})\vecb{F}_j\nonumber \\
  &+\sum_{j=1}^N \sum_\vecb{K\ne 0} M^{(2)}(\vecb{K})\cos(\vecb{K}\rij)
  \vecb{F}_j\nonumber \\ 
  &+\one\left(1-\frac{3}{2}\pi^{-1/2}a\alpha-
  \frac{2}{3}\pi^{-1/2}a^3\alpha^3\right)\vecb{F}_i
\end{align}
The prime in the first sum indicates, that for $\rij=0$ the terms with
$\vecb{n}=0$ are omitted. We used the definitions:
\begin{align}
  \label{eq34}
  M^{(1)}&(\vecb{r})=\one\left\{\left(\frac{3}{4}ar^{-1}+
  \frac{1}{2}a^3r^{-3}\right) {\rm erfc}(\alpha r) \right. \nonumber\\
  &+\left.a^3\alpha r^{-2}\pi^{-1/2} \exp(-\alpha^2 r^2)\right\} \nonumber\\
  &+\vecb{\hat{r}}\vecb{\hat{r}}\left\{\left(\frac{3}{4}ar^{-1}-
  \frac{3}{2}a^3r^{-3}\right) {\rm erfc}(\alpha r)
  +\left(-3a^3\alpha r^{-2}\right.\right.\nonumber\\
  &+\left.\left.\frac{3}{2}a\alpha-2a^3\alpha^3\right)
  \pi^{-1/2}\exp(-\alpha^2 r^2)\right\}
\end{align}
for the real part of the tensor and
\begin{align}
   \label{eq35}
  M^{(2)}&(\vecb{k})=\one\left\{2a\,{\rm erfc}
  \left(\frac{k}{2\alpha}\right)\right\}\frac{3\pi}{2L^2k}\nonumber\\
  &-\vecb{\hat{k}}\vecb{\hat{k}}\left\{\left(a-\frac{2}{3}a^3k^2\right) {\rm
    erfc}\left(\frac{k}{2\alpha}\right)\right.\nonumber\\ 
  &+\left.a\alpha^{-1}k\pi^{-1/2}
  \exp\left(-\frac{k^2}{4\alpha^2}\right)\right\}\frac{3\pi}{2L^2k} 
\end{align}
for the summation in Fourier space. This completes the derivation of the
quasi two--dimensional Ewald sum of the Rotne--Prager--Yamakawa mobility tensor.

The main advantage of the Rotne--Prager--Yamakawa tensor compared to the Oseen
tensor is its positive definiteness~\cite{Rotne:1969}, needed for the Cholesky
decomposition as used in Brownian dynamics
simulations~\cite{Allen:1987}. Therefore, one also has to consider the part of
the Rotne--Prager--Yamakawa tensor for $r_{ij}<2a$ (Eq. (23) in
Ref.~\cite{Rotne:1969}). However, this term does not contribute to the
long--ranged part of the Hydrodynamic Interaction and thus can be simply taken
out from the real space lattice sum for $\vecb{n}=0$ in Eq. (\ref{eq32}) or
Eq. (\ref{eq33}) and added separately. For the sum over the periodic images,
this term will not appear since $|\rij+\vecb{n}| >2a$ is always fulfilled.

\section{Results from simulations}
\label{sec:3}
In order to test the Ewald summation in a quasi--2D system using the
procedure outlined above, we performed Brownian dynamics
simulations with a single layer of $N$ colloids (Radius $a$) within a
fluid phase in a box with side lengths $L_x=L_y=L$, $L_z$. This system
resembles a model for the setup used in the experiments of
refs.~\cite{Rinn:1999,Zahn:1997} where paramagnetic colloidal particles were
placed atop a flat and stabilized air--water interface of a suspended
droplet. The colloids are fully immersed in the fluid phase and, due to
gravity pulling them downwards, just stay in the vicinity of the
interface~\cite{Oettel:2008} \tb{without perturbing the latter
  considerably}. Thus the colloidal particles constitute a quasi
two--dimensional system. More details can be found in
refs.~\cite{Rinn:1999,Zahn:1997}. In the model, the layer of particles is placed
parallel to the $(x-y)$--plane at $z=L_z/2$, with only in--plane motion
allowed. The colloids interact through a
repulsive potential $v_d/\kt=\Gamma/d^3$, where $d$ is the distance between 
each two particles scaled by the mean interparticle separation:
$d=r/\sqrt{\varrho}$ ($\varrho$ denotes the 2D number density of the
colloids). In the experimental setup, this dipole repulsion is generated and
controlled by an external magnetic field applied perpendicular to the layer of
colloids.

\begin{figure}[ht!]
  \psfig{file=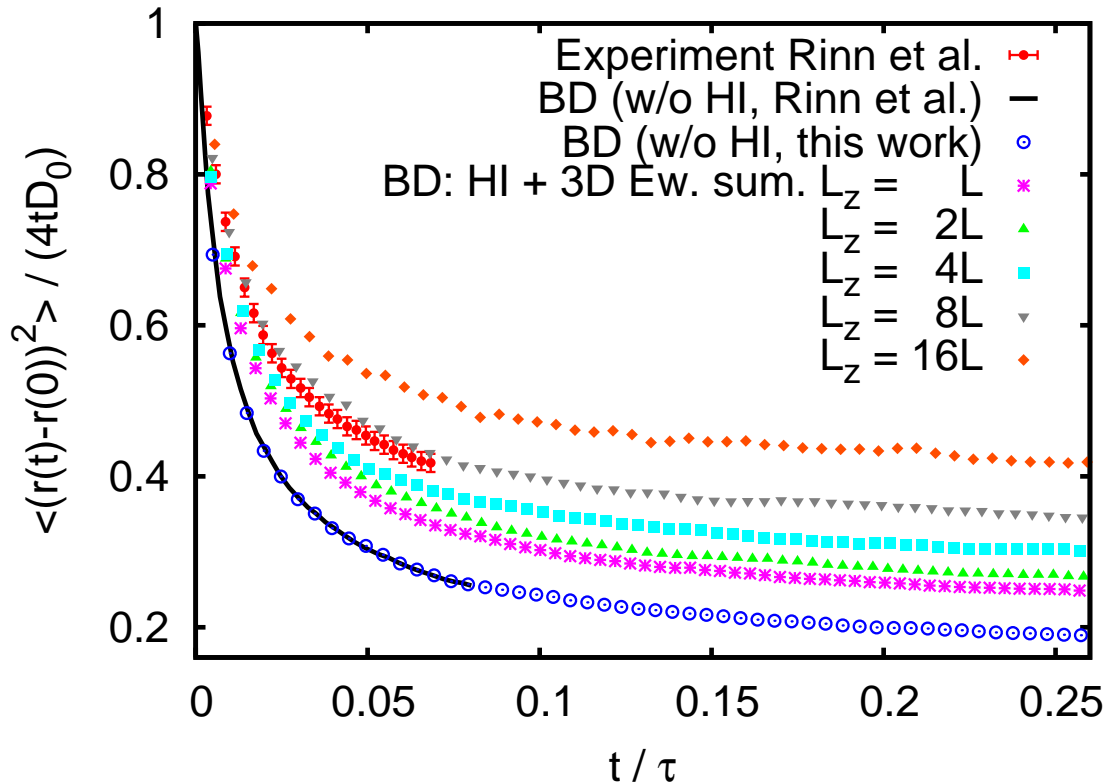,height=0.95\linewidth,angle=270}
  \caption{\label{fig1} Comparison of the scaled mean--squared displacement for
    \tb{Brownian dynamics simulations of} $N=100$ colloidal particles (radius
    $a=2.35 \mu$m, 2D number density $\varrho=3.24\cdot 10^{-3}\mu$m$^{-2}$)
    with a repulsion strength of $\Gamma=8.2$ \tb{without
      hydrodynamical interactions (open circles) and with HI included via 3D
      Ewald summation for a} varying longitudinal system size
    $L_z$. Experimental data \tb{and additional BD data (line)} were taken
    from ref.~\cite{Rinn:1999}. \tb{Errorbars are of the order of the
      corresponding symbol size and have been omitted for clarity}. A
    simulation with $L_z\approx 7L$ (not shown) would accidently match the
    data from the experiment. Such a fortuitous choice could well be the
    underlying reason for the good agreement between simulations with
    hydrodynamic interactions and the experimental data, as reported in
    ref.~\cite{Rinn:1999}.}   
\end{figure}

First, we validate our simulation by attempting to reproduce the Brownian
dynamics data published by Rinn et al.~\cite{Rinn:1999}. We therefore set
$\Gamma=8.2$ and $a=2.35 \mu$m for a system of $N=100$ colloids at a number
density of $\varrho=3.24\cdot 10^{-3}$\tb{, which corresponds to an area
  fraction of $\eta=0.056$}. The time is measured in units of 
$\tau=1/(\varrho D_0)$, where $D_0$ is the single particle short--time
diffusion constant, extracted from the simulations by extrapolating the
mean--squared displacement towards $t\to 0$~\cite{Rinn:1999}. In our
simulations we used the bulk value for the diffusion constant ($D=(6\pi\eta
a)^{-1}$). If the scaling of the mean--squared displacement  with $D$ holds,
the extrapolated short--time diffusion constant $D_0$ should agree with this
value. Indeed, for Brownian dynamics simulations we find $D_0=D$ and the
data agrees with ref.~\cite{Rinn:1999} (see Fig. \ref{fig1}). Furthermore,
we compare in Fig. \ref{fig1} the mean--squared displacement (scaled by
$4tD_0$) for simulations with varying system size $L_z$.
First it should be noted that the the extracted short time diffusion
constant $D_0$ differs from the corresponding bulk value $D$, if 3D Ewald
summation is applied. The extracted values increase with increasing
system size in $z-$direction. Additionally, as can be seen from the
simulation data presented in Fig. \ref{fig1}, that the mean squared
displacement, scaled by the corresponding short time diffusion constant
$D_0$, also increases with increasing system size. Thus the simulations
confirm the finding of a diverging mobility matrix for the 3D Ewald
summation of this system.

Neglecting noise, the 3D summation method remains valid. If we then
compare the effective velocities of single particles, we found that 
for our current setup, the distance of the periodic images in $z$-direction
had to be scaled by a factor of two \tb{relative to the original size
  $L_z=L_{x,y}$} in order to reduce the impact on the
single layer and to converge to the result from 2D Ewald summation according
to Eq. (\ref{eq32}). Note, that for large systems, such a stretching of the
third dimension could be unnecessary, since the size of a cubed box could
be already sufficient. However, this has to be checked for each setup
individually.  

As a second test, we perform simulations with hydrodynamical interactions
included, but without applying any Ewald summation procedure. We extract the
long--time self diffusion constant $D_L$ by running the simulation for much
longer times $t=1.2\tau$ and fitting the scaled mean--squared displacement for
times $t>0.95 \tau$ to a constant. 
\begin{figure}[ht!]
  \psfig{file=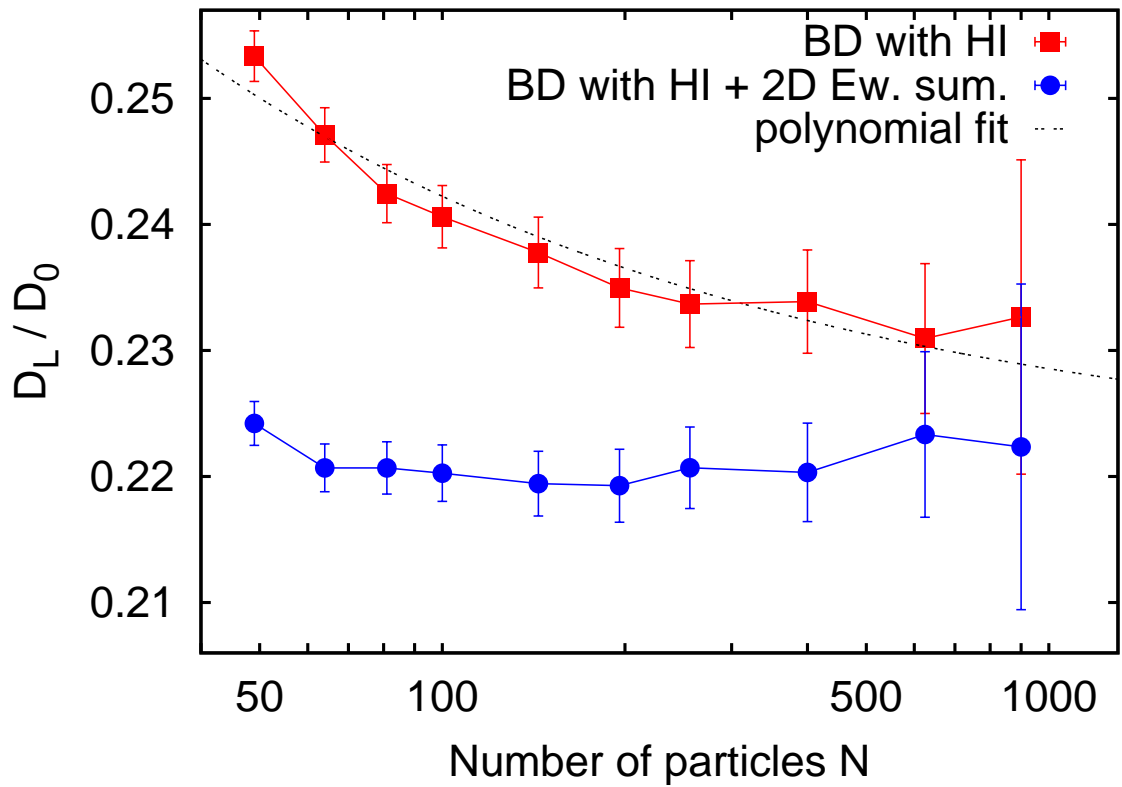,height=0.95\linewidth,angle=270}
  \caption{\label{fig2}Long--time self diffusion constant (scaled by $D_0$)
    extracted from simulations with hydrodynamical interactions,
    with and without 2D Ewald summation. As the number of particles in the
    system increases, $D_L$ approaches the value as obtained from simulations
    with 2D Ewald summation. Lines are drawn to guide the eyes, the dashed
    line corresponds to a polynomial fit. Error bars correspond to the
    statistical error obtained from averaging over many simulation runs.}  
\end{figure}
Fig. \ref{fig2} depicts the long--time diffusion constant (scaled by $D_0$) for
simulations with the same parameter setup as before and for various system
sizes and constant number density. As the number of particles increases,
for the Brownian dynamic simulation with HI, $D_L$ is found to decrease and
approach the limit set by using 2D Ewald summation. \tb{The system size
  dependence of $D_L$ without Ewald summation becomes clearly visible. In
  order to roughly characterize the convergence, we extrapolated} this
decrease using a polynomial fit for $0 < N < 700$. \tb{It turned out that} a
system containing $N \gtrsim 25000$ particles would yield a similar result for
$D_L/D_0$ as obtained in simulations with 2D Ewald summation. For the latter
ones, we do not observe any significant dependence of $D_L$ on the system size.

Of course it is now tempting to compare the 2D Ewald summation results to
the experimental data from ref.~\cite{Rinn:1999}. However, one has to keep in
mind, that the situation in the experiment is different. The particles are
adsorbed to a free interface, whereas there is none in the simulations. The
authors suggested to increase the \textit{hydrodynamical radius} in the
simulations, that is, the value of the particle radius within
the calculation of the mobility tensor was increased by a factor of two.
\begin{figure}[ht!]
  \psfig{file=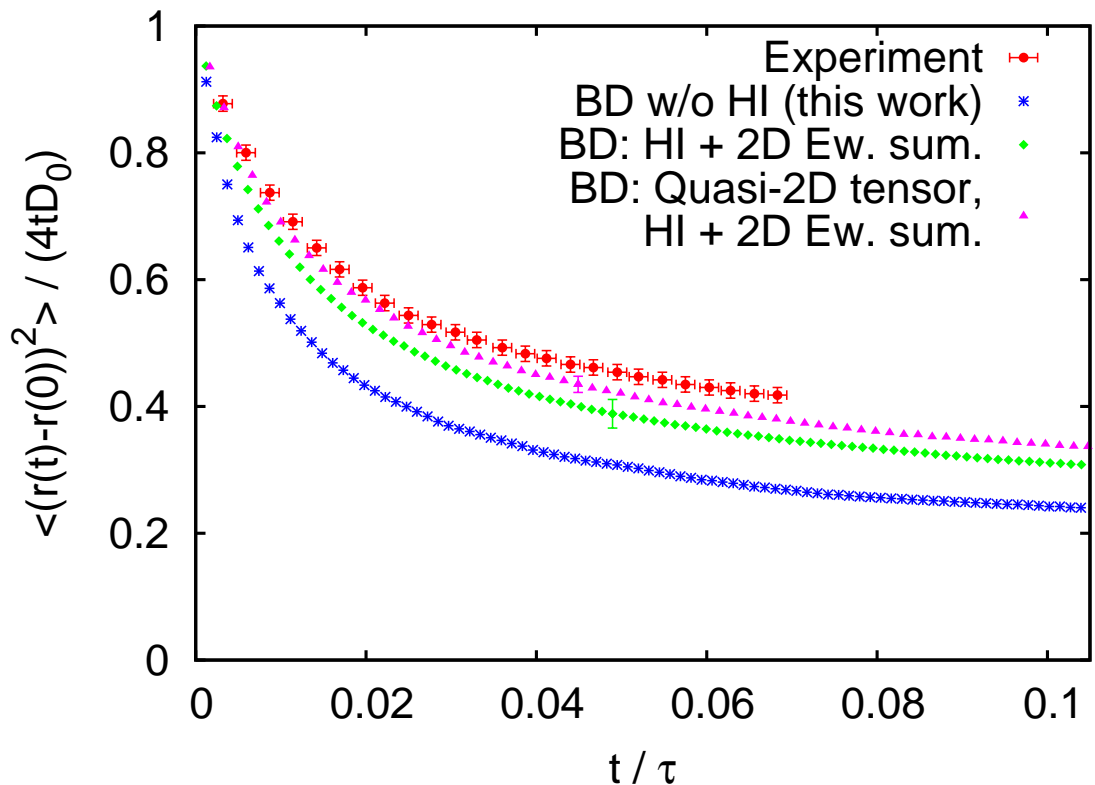,height=0.95\linewidth,angle=270}
  \caption{\label{fig3}Comparison of the simulated scaled mean--squared
    displacement for the same setup as in fig. \ref{fig1} to experimental data
    as obtained by the Experiment of Rinn et al.~\cite{Rinn:1999}. \tb{Only
      representative error bars are shown for simulations. The error bars for
      the Brownian dynamics simulation without hydrodynamical interactions
      (stars), are smaller than the symbol size.} The  
    simulation data stemming from \tb{BD simulations with
      hydrodynamical interactions and} 2D Ewald summation of the quasi
    2D mobility tensor \tb{of Cichocki et al.~\cite{Cichocki:2004}}
    (triangles) has been scaled by their different short time 
    diffusion constant $D_0=1.38D$.} 
\end{figure}
Fig. \ref{fig3} depicts the scaled mean--squared displacement for the 2D
Ewald summation method and the experimental data of
ref.~\cite{Rinn:1999}. The data cannot be reproduced with this actual
implementation. We refrained from trying out various hydrodynamical radii,
since this would introduce a free parameter to the system. 
If we use the mobility tensor by Cichocki et al.~\cite{Cichocki:2004}\tb{,
  derived from the two--sphere mobility tensor for particles close to a free
  interface as an asymptotic series in powers of $1/R$}  
(c. f. \ref{appc}, Eqs. (\ref{eqCI1}) and (\ref{eqCI1b})), the simulation
data overestimate the experimental values. However, the short time diffusion
constant $D_0$ extracted from the simulation is also increased, compared to
the simulations based on the Rotne--Prager mobility tensor. If we normalize
the quasi 2D simulation data with respect to the extracted value for $D_0$,
we find a rather good agreement. For longer times, however, the simulation
data starts to deviate slightly. Note that the deviation of the extrapolated
short time diffusion constant $D_0$ compared to $D=(6\pi\eta a)^{-1}$ may be
anticipated by inspection of Eqs. (\ref{eqCI1}) and (\ref{eqCI1b}) in the
appendix, since the matrix $\vecb{Q}_1$ for the self--diffusion already
deviates by a factor of $\approx 1.38$ from the unity matrix in the
Rotne--Prager case. Therefore, this deviation is only present in the quasi
2D case, for all other simulations the value of $D_0$ agrees with the
diffusion constant $D$ initially plugged in.

Concerning the computational cost of the simulations, we found that using the
2D Ewald summation method lead to a reduction of at least an order of magnitude,
compared to its 3D version, and depending on the particular choice of the
convergence parameter $\alpha$. For a typical choice of $\alpha=2/L$ the gain
in speed was around a factor $\sim 15$. Note that the computational cost of
the Ewald summation may vary, depending on the choice of $\alpha$.

\section{Summary and conclusions} 
\label{sec:4}
In summary, we have provided Ewald summation formulae for quasi two--dimensional
systems for the Rotne--Prager--Yamakawa mobility tensor and variants. We
demonstrated, that for quasi--two--dimensional systems, 3D Ewald summation
leads to a spurious system size dependence, stemming from the summation in the
direction perpendicular to the 2D layer of particles. This problem was
solved by summing in two dimensions only, and, additionally, using the
resulting formulae to calculate hydrodynamic interactions in computer
simulations of quasi 2D systems was found to be much more efficient, due to
the avoidance of summation in the third direction. We further found that the
asymptotic value of the long time diffusion constant for large systems could
already be obtained for rather small systems using the 2D Ewald summation
procedure. We demonstrated, that the 2D Ewald sum of the
quasi--two--dimensional analog of the Rotne--Prager mobility tensor given by
Cichocki and collaborators may be used to reproduce experimental data quite
well . Together with recent advances for an approximate and efficient
treatment of HI in computer simulations~\cite{Winter:2008}, inclusion of HI
and the proper treatment of their long--ranged characteristic becomes
feasible even for large systems in quasi 2D simulations of colloidal
suspensions.  

\label{acknowl}

J.B. thanks M. Oettel for fruitful discussions and the German Research
Foundation (DFG) for the financial support through the Collaborative Research
Center (SFB-TR6) ``Colloids in External Fields'' Project N01. 


\appendix
\section{Gradient terms and Integrals}

Here we list the evaluation of the gradients
$\nabla_{\vecbm{\xi}}\nabla_{\vecbm{\xi}}$ for the integrals appearing in the
Equations for $\Theta$ and $\chi$ (Eqs. (\ref{eq25}-\ref{eq28})), as well as
the resulting integrals that need to be performed in order to obtain
Eq. (\ref{eq32}):  
\begin{align}
  \label{A1}
  \nabla_{\vecbm{\xi}}\nabla_{\vecbm{\xi}}&\left.\int_0^{\alpha^2} t^{-1/2}
  \exp(-\frac{|\vecb{K}+\vecbm{\xi}|^2}{4t}-z^2t)\, dt 
  \,\right|_{\vecbm{\xi}=0} =\nonumber\\ 
  &-\frac{\one}{2}\int_0^{\alpha^2} t^{-3/2}
  \exp(-\frac{\vecb{K}^2}{4t}-z^2t)\, dt \nonumber\\
  &+\frac{\vecb{K}\vecb{K}}{4} \int_0^{\alpha^2} t^{-5/2}
  \exp(-\frac{\vecb{K}^2}{4t}-z^2t)\, dt
\end{align}
\begin{align}
  \label{A2}
  \nabla_{\vecbm{\xi}}\nabla_{\vecbm{\xi}}&\left.\int_0^{\alpha^2} t^{1/2}
  \exp(-\frac{|\vecb{K}+\vecbm{\xi}|^2}{4t}-z^2t)\, dt 
  \,\right|_{\vecbm{\xi}=0} =\nonumber\\ 
  &-\frac{\one}{2}\int_0^{\alpha^2} t^{-1/2}
  \exp(-\frac{\vecb{K}^2}{4t}-z^2t)\, dt \nonumber\\
  &+\frac{\vecb{K}\vecb{K}}{4} \int_0^{\alpha^2} t^{-3/2}
  \exp(-\frac{\vecb{K}^2}{4t}-z^2t)\, dt
\end{align}
The integral containing $t^{-3/2}$ has already been calculated (see
Eq. (\ref{eq18})), the remaining integrals can also be performed:
\begin{align}
  \label{A3}
  &\int_0^{\alpha^2} t^{-1/2}  \exp(-\frac{K^2}{4t}-z^2t) dt = \nonumber \\
  &\quad\frac{\sqrt{\pi}}{2z}\left[e^{-Kz}{\rm erfc}(\frac{K}{2\alpha}-\alpha
    z) +e^{Kz}{\rm erfc}(\frac{K}{2\alpha}+\alpha z)\right]\\
  \label{A4}
  &\int_0^{\alpha^2} t^{-5/2} \exp(-\frac{K^2}{4t}-z^2t) dt = \nonumber \\
  &\quad\quad\quad\frac{\sqrt{2\pi}}{K^3}\left[(1+Kz)e^{-Kz}{\rm
        erfc}(\frac{K}{2\alpha}-\alpha z) \right.\nonumber\\ 
  &\quad\quad\quad+\left.(1-Kz)e^{Kz}{\rm erfc}(\frac{K}{2\alpha}+\alpha
    z)\right]\nonumber\\
  &\quad\quad\quad+\frac{4}{\alpha K^2}\exp(-\frac{K^2}{4\alpha^2}-z^2\alpha^2)
\end{align}  
Taking the limit $z \to 0$ is straightforward for all terms except for the
integral in Eq. (\ref{A3}). This evaluates to: 
\begin{align}
  \label{A5}
  \lim_{z\to 0} \int_0^{\alpha^2}& t^{-1/2}  \exp(-\frac{K^2}{4t}-z^2t) dt =
  \nonumber \\ 
  &2\alpha \exp(-\frac{K^2}{4\alpha^2})-\sqrt{\pi}K{\rm erfc}(\frac{K}{2\alpha})
\end{align} 

\section{Ewald sum of the Oseen tensor}

For the Ewald sum according to Eq. (\ref{eq4}) of the Oseen tensor 
\addtocounter{equation}{+1}
\begin{align*}
  \label{eqB1}
  {\tag{\theequation a}}
  \vecb{O}_{ij} &=(8\pi\eta)^{-1} r^{-1}_{ij}(\one
  +\hrij\hrij)\,,\quad (i\ne j)\\   
  {\tag{\theequation b}}
  \vecb{O}_{ii}&=(6\pi\eta a)^{-1}\one\,,\quad (i=j)
\end{align*}
instead of $\MT_{ij}$, we neglect all terms in Eq. (\ref{eq32}) involving
$a^3$, since these are the additional terms of the Rotne--Prager--Yamakawa
tensor compared to the Oseen tensor. This leads to:
\begin{align}
  \label{eqB2}
  &(6\pi\eta a)\vecb{v}_{i,\rm eff}=\vecb{F}_i
  -\frac{3a\alpha}{2\sqrt{\pi}}\,\vecb{F}_i \nonumber\\
  &+\sum_{j=1}^N\left\{\sum_\vecb{n}\,^{\bm \prime}\,   
  \frac{3}{4}a\frac{{\rm erfc}(\alpha\R)}{\R}\left[\one+
    \hR\hR\right] \right.\nonumber\\
  &\left. +\frac{3}{2} \frac{a\alpha}{\sqrt{\pi}}
    e^{-\alpha^2\R^2}\hR\hR\right.\nonumber\\
  &+\left.\sum_{\vecb{K}\ne 0} \cos(\vecb{K}\rij)
  \frac{3a\pi}{L^2K} {\rm erfc}\left(\frac{K}{2\alpha}\right) \left[
    \one-\frac{1}{2}\vecb{\hat{K}}
    \vecb{\hat{K}}\right]\right.\nonumber\\  
  &-\left.\frac{3a\sqrt{\pi}}{2L^2\alpha} e^{-\frac{K^2}{4\alpha^2}}
      \vecb{\hat{K}}\vecb{\hat{K}}\right\}\vecb{F}_j      
\end{align}
Which also may be casted in a similar form as Eqs. (\ref{eq33}-\ref{eq35}):
\begin{align}
  \label{eqB3}
  (6\pi\eta a)\vecb{v}_{i,\rm eff}&=\sum_{j=1}^N \sum_\vecb{n}\,^{\bm \prime}
  M^{(1)}_O(\vecb{R}_{j,\vecb{n}})\vecb{F}_j\nonumber \\
  &+\sum_{j=1}^N \sum_\vecb{K\ne 0} M^{(2)}_O(\vecb{K})\cos(\vecb{K}\rij)
  \vecb{F}_j\nonumber \\ 
  &+\one\left(1-\frac{3}{2}\pi^{-1/2}a\alpha\right)\vecb{F}_i
\end{align}
with 
\begin{align}
  \label{eqB4}
  &M^{(1)}_O(\vecb{r})=\one\left\{\frac{3}{4}ar^{-1}
        {\rm erfc}(\alpha r)\right\} \nonumber\\
  &+\vecb{\hat{r}}\vecb{\hat{r}}\left\{\frac{3}{4}ar^{-1}
        {\rm erfc}(\alpha r)+ \frac{3}{2}a\alpha \pi^{-1/2}\exp(-\alpha^2
        r^2)\right\} 
\end{align}
for the real part of the tensor and
\begin{align}
  \label{eqB5}
  &M^{(2)}_O(\vecb{k})=\one\left\{2a\,{\rm erfc}
  \left(\frac{k}{2\alpha}\right)\right\}\frac{3\pi}{2L^2k}\nonumber\\
  &-\vecb{\hat{k}}\vecb{\hat{k}}\left\{a\, {\rm
    erfc}\left(\frac{k}{2\alpha}\right) 
  +a\alpha^{-1}k\pi^{-1/2}
  \exp\left(-\frac{k^2}{4\alpha^2}\right)\right\}\frac{3\pi}{2L^2k} 
\end{align}
for the Fourier space.

\section{Ewald sum for the quasi 2D mobility tensor of Cichocki et al.}
\label{appc}

The mobility tensor reported in Ref.~\cite{Cichocki:2004} for a quasi 2D
system of spherical particles close to a fluid interface differs from the 3D
Rotne--Prager--Yamakawa tensor by a factor of two for the terms proportional to
$1/r_{ij}$ and by a factor of $q=5.59027$ for terms $\propto
1/r_{ij}^3$. Additionally, it uses different matrices for the self mobility and 
for terms $\propto 1/r_{ij}^3$ instead of the unity matrix:  
\addtocounter{equation}{+1}
\begin{align*}
  \label{eqCI1} 
  \MT_{ij} &=(6\pi\eta a)^{-1}\left\{\frac{3}{2}ar^{-1}_{ij}(\one
  +\hrij\hrij)\right.\\ 
  {\tag{\theequation a}}
  &+\left.\frac{1}{2}a^3r^{-3}_{ij} q
  (\vecb{Q}_3-3\hrij\hrij)\right\}\,,\quad (i\ne 
  j)\\    
  {\tag{\theequation b}}
  \label{eqCI1b}
  \MT_{ii}&=(6\pi\eta a)^{-1} \vecb{Q}_1,\quad (i=j)
\end{align*}
with $q=5.59027$ and matrices $\vecb{Q}_1$ and
$\vecb{Q}_3$~\cite{Cichocki:2004}  
\begin{equation}
  \label{eqCI2}
  \vecb{Q}_1=\begin{pmatrix} 1.3799554 & 0\\0 & 1.3799554 \end{pmatrix}
\end{equation}
\begin{equation}
  \label{eqCI3}
  \vecb{Q}_3=\begin{pmatrix} -0.319658 & 0\\0 & -0.319658 \end{pmatrix}
\end{equation}
reflecting the presence of a free interface (the additional scaling factor
$q$ and the matrix $\vecb{Q}_3$ may be derived from 
Eq.(8) in Ref.~\cite{Cichocki:2004} by casting it into the Rotne--Prager form
(Eq.(\ref{eq3}))). As a consequence of that, the resulting summation
formula involves more terms, since the different matrices avert
cancellations. The corresponding analog to Eq.(\ref{eq32}) then reads: 
\begin{align}
  \label{eqCI4}
  &(6\pi\eta a)\vecb{v}_{i,\rm eff}=\vecb{Q}_1\vecb{F}_i
  -\frac{a\alpha}{\sqrt{\pi}}
  \left(3+\frac{2}{3}q a^2\alpha^2\vecb{Q}_3\right)\vecb{F}_i\nonumber\\
  &+\sum_{j=1}^N\left\{\sum_\vecb{n}\,^{\bm \prime}\,\left( \frac{3}{2}a\frac{{\rm
      erfc}(\alpha\R)}{\R}\left[\one+ \hR\hR\right] \right.\right.\nonumber\\
  &+\left.\left[\frac{1}{2}q a^3\frac{{\rm erfc}(\alpha\R)}{\R^3}
    +\frac{q a^3\alpha}{\sqrt{\pi}}
    \frac{e^{-\alpha^2\R^2}}{\R^2}\right] \left[\vecb{Q}_3-3\hR\hR\right]
  \right. \nonumber\\ 
  &\left. +\frac{a\alpha}{\sqrt{\pi}}
  e^{-\alpha^2\R^2} \left[3-2q a^2\alpha^2\right]\hR\hR\right)\nonumber\\
  &+\left.\sum_{\vecb{K}\ne 0} \cos(\vecb{K}\rij)\left(
  \frac{6a\pi}{L^2K} {\rm erfc}\left(\frac{K}{2\alpha}
  \right)\right.\right.\nonumber\\  
  &\times\left(\one-\frac{1}{2}\vecb{\hat{K}}
    \vecb{\hat{K}}+\frac{1}{6}q a^2K^2\left[\one-\vecb{Q}_3+\vecb{\hat{K}}
    \vecb{\hat{K}}\right]\right)\nonumber\\
  &\left.\left.  
  -\frac{3a\sqrt{\pi}}{L^2\alpha} e^{-\frac{K^2}{4\alpha^2}}
  \left[\vecb{\hat{K}}\vecb{\hat{K}}+\frac{2}{3}q a^2\alpha^2
    (\one-\vecb{Q}_3)   
    \right]\right)\right\}\vecb{F}_j    
\end{align} 
and accordingly
\begin{align}
  \label{eqCI5}
  (6\pi\eta a)&\vecb{v}_{i,\rm eff}=\sum_{j=1}^N \sum_\vecb{n}\,^{\bm \prime}
  M^{(1)}_{q2D}(\vecb{R}_{j,\vecb{n}})\vecb{F}_j\nonumber \\
  &+\sum_{j=1}^N \sum_\vecb{K\ne 0} M^{(2)}_{q2D}(\vecb{K})\cos(\vecb{K}\rij)
  \vecb{F}_j\nonumber \\ 
  &+\left(\vecb{Q}_1-3\pi^{-1/2}a\alpha-
  \frac{2}{3}\pi^{-1/2}q a^3\alpha^3\vecb{Q}_3\right)\vecb{F}_i
\end{align}
with the definitions:
\begin{align}
  \label{eqCI6}
  M^{(1)}_{q2D}&(\vecb{r})=\one\left\{\left(\frac{3}{2}ar^{-1}+
  \frac{1}{2}q a^3r^{-3}\vecb{Q}_3\right) {\rm erfc}(\alpha
  r)\right.\nonumber\\  
  &+\left.q a^3\alpha r^{-2}\pi^{-1/2} \exp(-\alpha^2 r^2)\vecb{Q}_3\right\} 
    \nonumber\\ 
  &+\vecb{\hat{r}}\vecb{\hat{r}}\left\{\left(\frac{3}{2}ar^{-1}-
  \frac{3}{2}q a^3r^{-3}\right) {\rm erfc}(\alpha r)
  +\left(-3q a^3\alpha r^{-2}\right.\right.\nonumber\\
  &+\left.\left.3 a\alpha-2q a^3\alpha^3\right)
  \pi^{-1/2}\exp(-\alpha^2 r^2)\right\}
\end{align}
for the real part of the tensor and
\begin{align}
   \label{eqCI7}
   M^{(2)}_{q2D}&(\vecb{k})=\one\left\{\left(2a+\frac{1}{3}q a^3k^2
   (\one-\vecb{Q}_3)\right){\rm erfc}\right.
   \left(\frac{k}{2\alpha}\right)\nonumber\\
   &-\frac{2}{3}q a^3\alpha \left.
   k\pi^{-1/2}\exp\left(-\frac{k^2}{4\alpha^2}\right)(\one-\vecb{Q}_3)
   \right\}\frac{3\pi}{L^2k} \nonumber\\
   &-\vecb{\hat{k}}\vecb{\hat{k}}
   \left\{\left(a-\frac{1}{3}q a^3k^2\right) {\rm erfc}
   \left(\frac{k}{2\alpha}\right)\right.\nonumber\\  
   &+\left.a\alpha^{-1}k\pi^{-1/2}
   \exp\left(-\frac{k^2}{4\alpha^2}\right)\right\}\frac{3\pi}{L^2k} 
\end{align}
in Fourier space.

\section{Ewald sum for the binary Rotne--Prager Tensor}

For binary mixtures of particles with radii $a_i \in \{a_0,\,a_1\}$,
we replace the particles radius $a$ by $a_i$, and within the sum over
particles with radius $a_j$, each factor $a^3$ is replaced by
$\frac{a_i}{2}(a_i^2+a_j^2)$~\cite{Garcia:1977}:  
\addtocounter{equation}{+1}
\begin{align*}
  \label{eqC1}
  \MT_{ij} &=(6\pi\eta a_i)^{-1}\left\{\frac{3}{4}a_ir^{-1}_{ij}(\one
  +\hrij\hrij)\right.\\ 
  {\tag{\theequation a}}
  &+\left.\frac{a_i}{4}(a_i^2+a_j^2)r^{-3}_{ij}({\bf
    1}-3\hrij\hrij)\right\}\,,\quad (i\ne j)\\   
  {\tag{\theequation b}}
  \MT_{ii}&=(6\pi\eta a_i)^{-1}\one\,,\quad (i=j)
\end{align*}
Carrying out this procedure for the previous results
(Eqs. (\ref{eq33}-\ref{eq35}) leads to:
\begin{align}
  \label{eqC2}
  (6\pi\eta a_i)&\vecb{v}_{i,\rm eff}=\sum_{j=1}^N \sum_\vecb{n}\,^{\bm \prime}
  M^{(1)}_b(\vecb{R}_{j,\vecb{n}})\vecb{F}_j\nonumber \\
  &+\sum_{j=1}^N \sum_\vecb{K\ne 0} M^{(2)}_b(\vecb{K})\cos(\vecb{K}\rij)
  \vecb{F}_j\nonumber \\ 
  &+\one\left(1-\frac{3}{2}\pi^{-1/2}a_i\alpha-
  \frac{2}{3}\pi^{-1/2}a_i^3\alpha^3\right)\vecb{F}_i
\end{align}
With the corresponding definitions:
\begin{align}
  \label{eqC3}
  M^{(1)}_b&(\vecb{r})=\one\left\{\left(\frac{3}{4}a_ir^{-1}+
  \frac{1}{4}a_i(a_i^2+a_j^2) r^{-3}\right) {\rm erfc}(\alpha r)
  \right. \nonumber\\ 
  &+\left.\frac{a_i}{2}(a_i^2+a_j^2) \alpha r^{-2}\pi^{-1/2} \exp(-\alpha^2
  r^2)\right\} \nonumber\\ 
  &+\vecb{\hat{r}}\vecb{\hat{r}}\left\{\left(\frac{3}{4}a_ir^{-1}-
  \frac{3}{4}a_i(a_i^2+a_j^2)r^{-3}\right) {\rm erfc}(\alpha r)
  \right.\nonumber\\ 
  &+\left(-\frac{3a_i}{2}(a_i^2+a_j^2)\alpha r^{-2}
  +\left.\frac{3}{2}a_i\alpha-a_i(a_i^2+a_j^2)\alpha^3\right)\right. \nonumber\\
  &\left.\times\,\pi^{-1/2}\exp(-\alpha^2 r^2)\right\}
\end{align}
for the real part of the tensor and
\begin{align}
   \label{eqC4}
  M^{(2)}_b&(\vecb{k})=\one\left\{2a_i\,{\rm erfc}
  \left(\frac{k}{2\alpha}\right)\right\}\frac{3\pi}{2L^2k}\nonumber\\
  &-\vecb{\hat{k}}\vecb{\hat{k}}\left\{\left(a_i-\frac{1}{3}a_i(a_i^2+a_j^2)
  k^2\right) {\rm erfc}\left(\frac{k}{2\alpha}\right)\right.\nonumber\\ 
  &+\left.a_i\alpha^{-1}k\pi^{-1/2}
  \exp\left(-\frac{k^2}{4\alpha^2}\right)\right\}\frac{3\pi}{2L^2k} 
\end{align}
for the summation in Fourier space.



\end{document}